\documentclass[twocolumn,trackchanges]{aastex63}
\usepackage{amsmath}
\usepackage{graphicx}
\usepackage{amsfonts}
\usepackage{natbib}
\usepackage{color}
\usepackage{epstopdf}
\usepackage{graphicx}
\usepackage{multirow}
\usepackage{longtable}
\usepackage{rotating}

\shorttitle{QPO in the blazar PKS 0405-385}
\shortauthors{Gong et al.}

\newcommand\gr{{$\gamma$-ray}}

\begin{document}

\title{Quasi-periodic behaviour in the $\gamma$-ray light curve of the blazar PKS 0405-385}

\author{Yunlu Gong}
\affil{Department of Astronomy, School of Physics and Astronomy, Key Laboratory of Astroparticle Physics of Yunnan Province, Yunnan University, Kunming 650091, People's Republic of China; fangjun@ynu.edu.cn, gongyunlu@qq.com}

\author{Liancheng Zhou}
\affil{Department of Astronomy, School of Physics and Astronomy, Key Laboratory of Astroparticle Physics of Yunnan Province, Yunnan University, Kunming 650091, People's Republic of China; fangjun@ynu.edu.cn, gongyunlu@qq.com}

\author{Min Yuan}
\affil{Department of Astronomy, School of Physics and Astronomy, Key Laboratory of Astroparticle Physics of Yunnan Province, Yunnan University, Kunming 650091, People's Republic of China; fangjun@ynu.edu.cn, gongyunlu@qq.com}

\author{Haiyun Zhang}
\affil{Department of Astronomy, School of Physics and Astronomy, Key Laboratory of Astroparticle Physics of Yunnan Province, Yunnan University, Kunming 650091, People's Republic of China; fangjun@ynu.edu.cn, gongyunlu@qq.com}

\author{Tingfeng Yi}
\affil{Physics Department, Yunan Normal University, Kunming 650092, People's Republic of China}

\author{Jun Fang}
\affil{Department of Astronomy, School of Physics and Astronomy, Key Laboratory of Astroparticle Physics of Yunnan Province, Yunnan University, Kunming 650091, People's Republic of China; fangjun@ynu.edu.cn, gongyunlu@qq.com}

\begin{abstract}

We analyze the quasi-periodic oscillation (QPO) of the historical light curve
of FSRQs PKS 0405-385 detected by the Fermi LAT from August 2008 to November 2021.
To identify and determine the QPO signal of PKS 0405-385 in the $\gamma$-ray light curve, we use four time
series analysis techniques based on frequency and time domains, i.e., the Lomb-Scargle periodogram
 (LSP), the weighted wavelet z-transform (WWZ), the REDFIT and the epoch folding. The results show that PKS 0405-385
 has a quasi-periodic behavior of $\sim$2.8 yr with the significance of $\sim$4.3$\sigma$ in Fermi long-term monitoring.
Remarkably, we also performed QPO analysis in the G-band light curve observed from October 2014 to October 2021
using LSP and WWZ technology, and the results ($\sim$4$\sigma$ of significance) are consistent with the periodic detection in $\gamma$-ray.
This may imply that the optical emission is radiated by an electron population same as the \gr\ emission.
In discussing the possible mechanism of quasi-periodic behavior, either the helical motion within a jet or
the supermassive black hole binary system provides a viable explanation for the QPO of 2.8 yr, and the relevant parameters have been estimated.
\end{abstract}

\keywords{galaxies: active - galaxies: individual: PKS 0405-385 - quasi-periodic oscillation}

\section{Introduction}
\label{sec:intro}

Blazars are a particular subclass of active galactic nuclei (AGN) whose relativistic jets
point near the observer's line of sight \citep{1993ARA&A..31..473A,1995PASP..107..803U}.
According to the presence or absence of emission lines in the optical/infrared spectrum,
blazars are divided into two subclasses, BL Lac Objects (BL Lacs, either weak or not)
and flat-spectrum radio quasars (FSRQs, evidently). In addition, a method for
physically distinguishing the two subclasses is proposed; that is, FSRQs is the ratio of the luminosity
of the broad-line region (BLR) to the Eddington luminosity greater than 5 $\times$ $10^{-5}$ \citep{2011MNRAS.414.2674G}.
Generally, blazars' spectral energy distributions (SEDs) present a double-peaked shape. The low energy component
(optical to soft X-ray wavelength) in the broadband SED of blazars is generated by synchrotron radiation from electrons in
the jet, while the high-energy component (hard X-ray to $\gamma$-ray wavelength) is produced by the inverse Compton scattering off
soft photons or by hadronic processes \citep{1974ApJ...188..353J,1992A&A...253L..21M,1993ApJ...416..458D}.

As one of the characteristics of blazars, rapid and violent variabilities across the entire
electromagnetic spectrum enable us to study the nature of blazars by analyzing the light curve. Quasi-periodic variability is one of the blazar's light curve characteristics.
Furthermore, studies of the QPOs of blazars allow us to explore the central engine's structures, physical properties, dynamics, and radiation mechanisms,
although QPOs are rare and transient in their multiwavelength light curves \citep{2021MNRAS.506.1540L}. When searching for sources with QPO behavior, false periodic events may occur if the number of cycles is too small. Therefore, the QPO examples introduced below are referenced from more to less according to the number of cycles observed.

In the radio domain, a QPO signal ($\sim$176 days) with 21 cycles was announced in J0849+5108 \citep{2021ApJ...914....1Z}.
In the same year, \citet{2021MNRAS.506.1540L} found about 850 days of periodic modulation in all three bands (4.8, 8, and 14.5 GHz) of the BL Lac OT 081,
and a pure geometric scenario provided a plausible explanation for the detected QPO.
Interestingly, a QPO of $\sim$4.69 yr ($>$5$\sigma$ confidence level) was found in PKS J2134-0153,
in which the 15GHz light curve is very close to sinusoidal variation \citep{2021MNRAS.506.3791R}. Then some possible QPOs have also been studied in J1359+4011,
PKS 0219-164, J1043+2408, and PKS J0805-0111 \citep{King2013,Bhatta2017,Bhatta2018,2021RAA....21...75R}.

In the optical frequencies, the widely studied case is OJ 287, which
reveals regular optical outbursts on a 12-year timescale that has been modeled as the result of a secondary SMBH companion
passing through the primary SMBH's accretion disk \citep{Kidger1992,Valtonen2006,Valtonen2008,2015ApJ...803L..16L}. Moreover,
many authors have found QPO with different timescales in the optical frequency of this source \citep{2013MNRAS.434.3122P,2016ApJ...832...47B}.
In addition, the quasar PG 1302-102 was observed by the Catalina Real-time Transient Survey (CRTS) and found a
quasi-periodic variation of $\sim$1884 days in the V band light curve \citep{2015Natur.518...74G}. Then some other
analogous candidates were also studied \citep{2015MNRAS.453.1562G}.
Other cases with possible QPO are also investigated, such as PKS 2155-304, 3C 279, SDSS J0159+0105, and PSO J334.2028+01.4075 \citep{2014RAA....14..933Z,Sandrinelli2014,2016AJ....151...54S,2015ApJ...803L..16L,2016ApJ...827...56Z}.

Interesting quasi-periodic emission phenomena can also be found in X-ray emission sources.
\citet{Gierlinski2008} revealed that a 1-h modulation with $\sim$23 cycles existed in RE J1034+396.
After that, more and more QPOs have been discovered in X-rays, which include Mrk 766 \citep[$\sim$1.8 hr,][]{Zhang2017d}, MCG-06-30-15 \citep[$\sim$1 hr,][]{2018A&A...616L...6G}, MS 2254.9-3712 \citep[$\sim$2.0 hr,][]{2015MNRAS.449..467A}, 1H 0707-495 \citep[$\sim$2.3 hr,][]{Zhang2018}, and 2XMM J123103.2+110648 \citep[$\sim$3.8 hr,][]{Lin2013}. Such a short QPO timescale has attracted extensive attention. This behavior may be related to the accretion of the innermost stable circular orbit around the black hole \citep{Kluzniak2002,Remillard2006,Zhang2017d,Zhang2018,2022MNRAS.509.3504B}.

Since the operation of the Fermi Gamma-ray Space Telescope in 2008, several cases of AGN QPOs have been reported in the $\gamma$-ray light curve.
The most well-known case is PKS 2247-131, which presents a relatively short periodicity in its $\gamma$-ray light curve from November 2016 to
June 2017, and the $\sim$34.5 day oscillation with six cycles is explained in terms of a helical structure in the jet \citep{Zhou2018}. Moreover, QPOs behaviors
of OJ 287 ($\sim$314 days), PKS 1424-418 ($\sim$355 days), PKS 0521-36 ($\sim$1.1 yr), PKS 2155-304 ($\sim$1.7 yr), PKS 0301-243 ($\sim$2.1 yr), and PKS 0426-380 ($\sim$3.35 yr) have been claimed by \citet{Sandrinelli2014}, \citet{Zhang2017a,Zhang2017b,Zhang2017c}, \citet{2020MNRAS.499..653K}, \citet{2021PASP..133b4101Y}, and \citet{2021ApJ...919...58Z}.
In multiwavelength bands, the similarity of low- and high- energy periodic modulation ($\sim$2 yr) in PG 1553+113 has been reported by \citet{2015ApJ...813L..41A}.
The possible QPOs of Mrk 421, BL Lacertae, 3C 454.3, CTA 102, PMN J0948+0022, S5 0716+714, and J0112.1+2245 in multi-band are also widely studied, although most of the modulation
are not similar at different frequencies \citep{2016PASP..128g4101L,Zhangjin,2017A&A...600A.132S,2020A&A...642A.129S,2021MNRAS.501...50S,2022Ap&SS.367....6G}.

PKS 0405-385 (also known as 4FGL J0407.0-3826) is identified as an FSRQ based on its robust and broad emission lines in the optical spectrum with a redshift $z$ = 1.285
\citep{1990A&AS...86..543V,1997ApJ...490L...9K}. Its observations commenced in November 1993, as part of the Australia Telescope Compact Array intraday variability (IDV) survey.
The source exhibits fluctuations on times of a day or less in the flux density at gigahertz frequency. The interstellar scintillations model can well explain such an IDV behavior, and then the value of the brightness temperature is near $10^{13}$K at 5 GHz \citep{2003MNRAS.341..230P}. \citet{2002ApJ...581..103R} inferred a Doppler factor of about 75 based on the IDV behavior in the radio band, which is greater than the Doppler factor from other AGN monitoring. After eleven years, \citet{2013RAA....13..259F} compiled the available $\gamma$-ray data from Fermi-LAT, and then a Doppler factor of 8.93 was evaluated. In the G-band, the Gaia satellite promulgated the first data point for the source with a magnitude of M = 18.15 in October 2014 and has been continuously observed until now \citep{2016A&A...595A...2G}. Nevertheless, Fermi-LAT also detected it as part of a family of high energies emitters with a hard photon index of $\Gamma$ = 2.40 $\pm$ 0.08 \citep{2010ApJ...715..429A}.

In this paper, based on the recent report on monitoring PKS 0405-385 with the Fermi Gamma-ray Space Telescope,
the investigation of the variability behavior of the source in \gr\ energies is carried out. Since 2014, the almost simultaneous
optical data of the source has been collected by Gaia, so we also investigate the variability behavior in the G band.
This paper is organized as follows. The analysis of the Gaia optical light curve and \emph{Fermi} \gr\ data is given in Section~\ref{data}.
The results of a periodic search of light curves in different bands by different methods are presented in Section \ref{results}.
The main conclusions and some discussion are given in Section \ref{sumdis}.

\section{Data Reduction and Analysis}
\label{data}
\subsection{Gaia light curve data}
The Gaia satellite was launched by the European Space Agency at the
end of 2013 and began scientific mission monitoring in July 2014. The Gaia satellite will allow the determination of highly accurate positions, parallaxes, and proper motions for $>$ 1 billion sources brighter than magnitude 20.7 in the white-light photometric band G of Gaia.
The survey significantly impacts a broad range of fields, such as cosmological gravitational lensing, white dwarfs,
and hypervelocity stars \citep{2020MNRAS.493.3264K,2021A&A...652A..76H}. The source PKS 0405-385 (R.A. = $04^h06^m59^s.040$,
Decl. = $-38^{\circ}26^{'}28^{''}.030$) was named Gaia18eai in the Gaia Photometric Science Alerts\footnote{http://gsaweb.ast.cam.ac.uk/alerts/alertsindex}.
The magnitude error can be obtained using a historical standard deviation of 0.51.
We present an optical light curve similar to the varies of \gr\ flux density in panel (a) of Figure~\ref{Fig4}.
By observing the light curve, it can be found that there are prominent flares in March 2016 and January 2019, respectively.
The data acquisition interval of the source is uneven due to the limitation of observation conditions, and the interval
range is from a few days to 72 days. During the Gaia satellite observation, the G-band light curve varies between 15.27 and
18.85, with a mean value of 17.82 and a standard deviation of 61.4\%.

\subsection{High-energy \gr\ : Fermi-LAT Data}
The Large Area Telescope (LAT) on the Fermi Gamma-ray Space Telescope was launched by NASA in June 2008.
Fermi LAT is designed to measure the directions, energies, and arrival times of \gr\ incident over a wide field of view, while rejecting background from cosmic rays \citep{Atwood2009}. Moreover, the LAT can monitor the all-sky every 90 minutes and detect photon events below 20 MeV to more than 300 GeV energies.

\begin{figure*}
\centering
\includegraphics[width=0.65\textwidth]{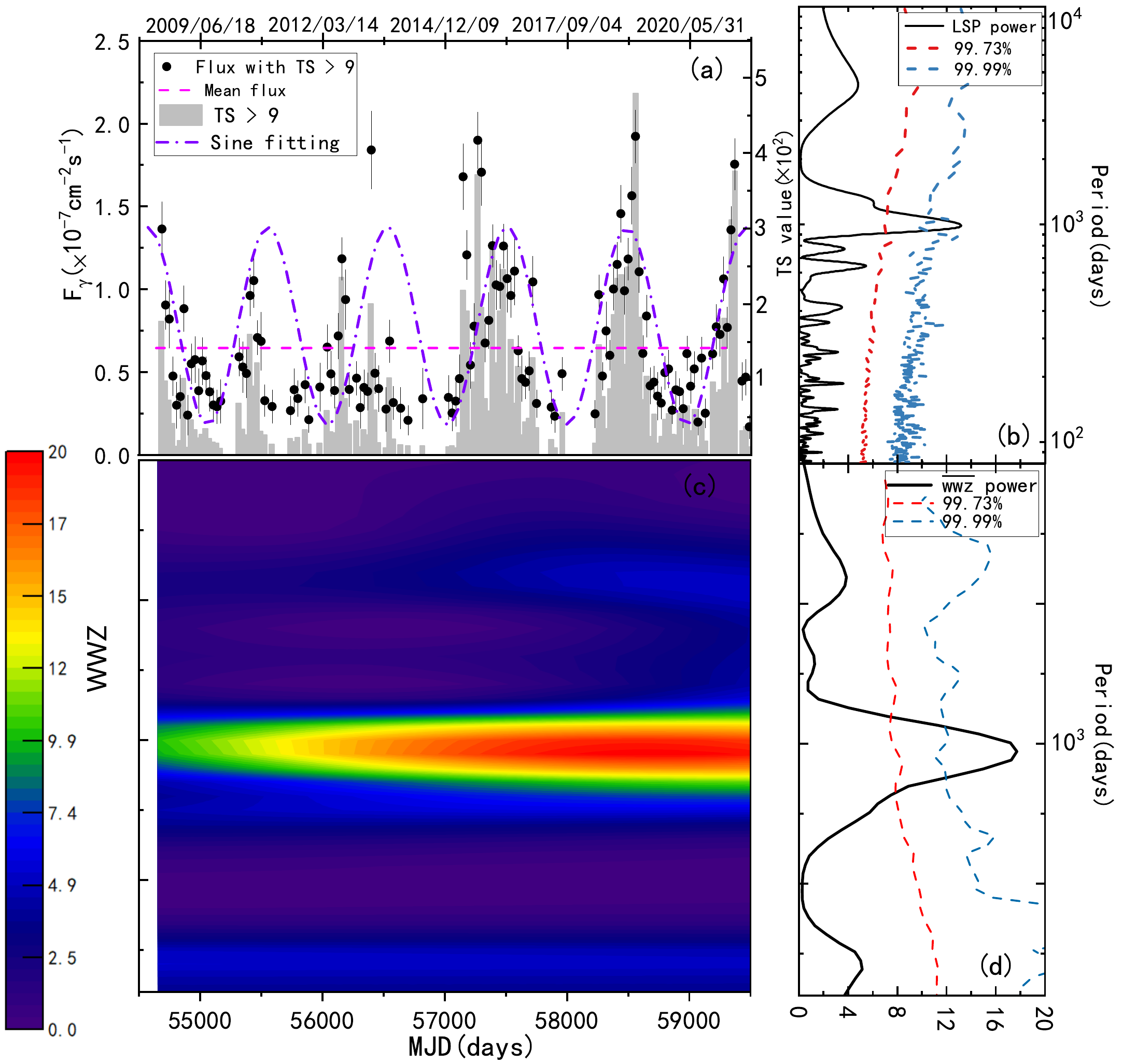}
%%\vspace{-20pt}\includegraphics[width=0.35\textwidth]{Fig3.pdf}
%%\plottwo{Fig1.pdf}{Fig3.pdf}
\caption{Panel (a): Light curves of the FSRQs PKS 0405-385 in \gr\ energy band with $TS\geqslant9$.
   The corresponding TS value of each data point is represented by gray histogram and
   the magenta horizontal dashed line indicates the mean flux density of the light curve. The purple dash-dotted line represents the sinusoidal fitting curve of the observation data.
   Panel (b): LSP power spectrum for the monthly binned \gr\ light curve (black solid line),
   and the red and blue dashed lines represent 99.73\% (3$\sigma$) and 99.99\% (4$\sigma$) significance contours, respectively.
   Panel (c): two-dimensional contour map of the WWZ power spectrum of the whole light curve.
   Panel (d): time-averaged WWZ power spectrum (black solid line) calculated from the data,
   3$\sigma$ and 4$\sigma$ significance contours (red and blue dashed lines) from artificial light curves.}
\label{Fig1}
\end{figure*}

To extract the light curve, we downloaded the Fermi-LAT\footnote{https://fermi.gsfc.nasa.gov/ssc/data/access/lat/} data processed with the Pass 8 instrument
response function for 4FGL J0407.0-3826 for the period between 2008 October 4 and 2021 November 2 (MET:239557417--657526922, $\sim$13.2 yr).
The result of a large point spread function in the low energies ($<$100 MeV) is likely to be unreliable, so the energy range from 100 MeV to 300 GeV is selected.
In these energy ranges, we selected the `SOURCE' class registered events from a $12^{\circ}$ circular region of interest centred on the source location (R.A. = $61^{\circ}.7627$, Decl. = $-38^{\circ}.4394$). In order to avoid photon confusion from the Earth's limb, a maximum zenith angle is limited to $90^{\circ}$.
A Good Time Interval is selected by using the expression $``(~DATA\_QUAL > 0)\&\&(~LAT\_CONFIG=1)"$.
The input XML model file contains two components: Galactic (gll\_iem\_v07) and isotropic extra-galactic (iso\_P8R3\_SOURCE\_V2\_v1.txt).
Through the above analysis, we get the integrated photon flux of $\rm (10.0\pm0.2)\times10^{-8}~photons\,cm^{-2}\,s^{-1}$ with a test statistic (TS) value of $\sim$ 8152.
Finally, we constructed a monthly (30 day bin) light curve with TS $\geqslant$ 9 ($\gtrsim 3\sigma$) to describe the variation trend of the source, as shown in panel (a) of Figure~\ref{Fig1}. The TS values of data points are presented in Figure~\ref{Fig1} with a gray histogram.

\begin{figure}
\centering
\includegraphics[scale=0.4]{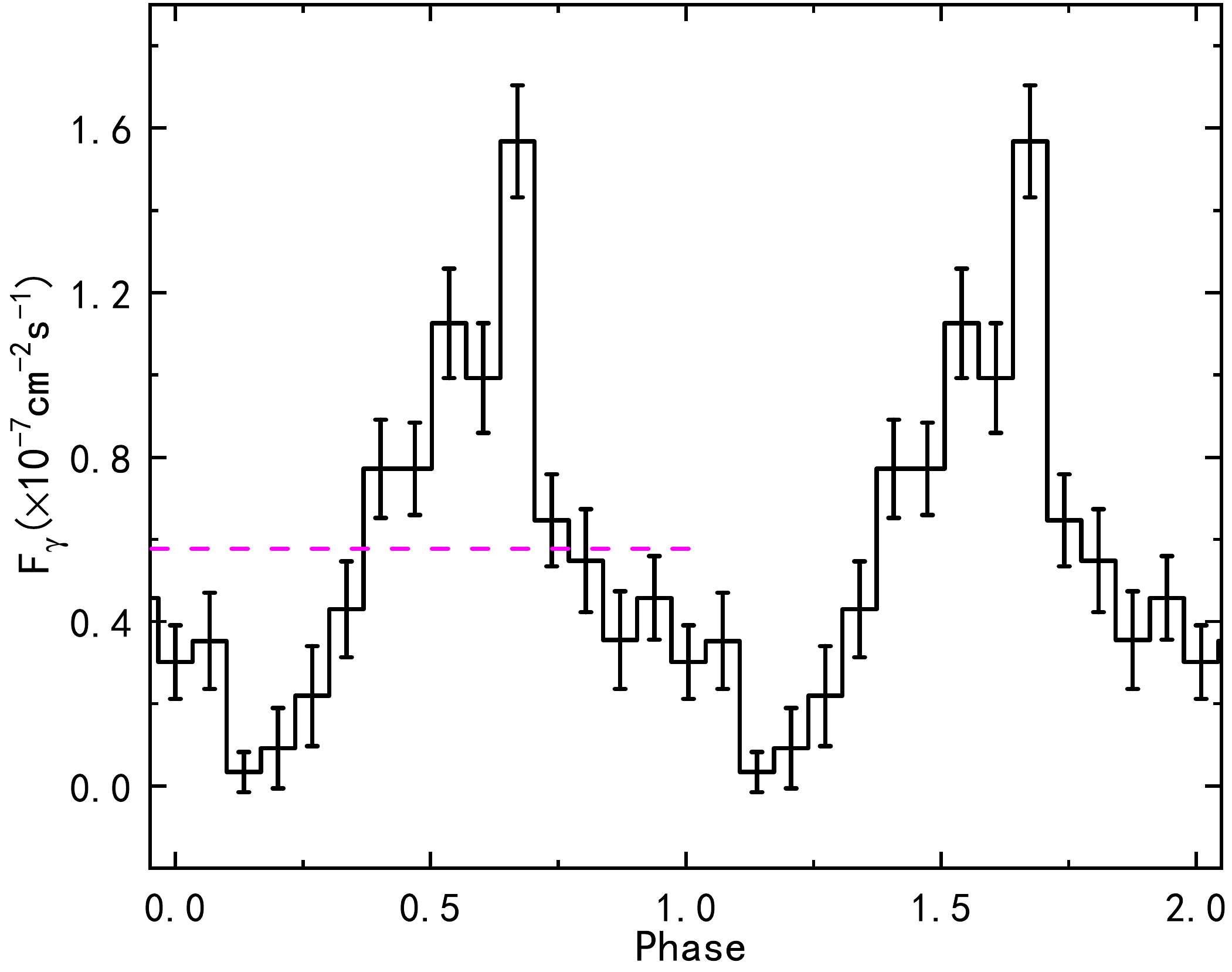}
\caption{Epoch-folded pulse shape from binned likelihood analysis of the
flux density with a period of 1022 days. The magenta dashed line represents the mean flux in the \gr\ light curve.
For clarity, we show two period cycles.}
\label{Fig2}
\end{figure}

\begin{figure}
\centering
\includegraphics[scale=0.4]{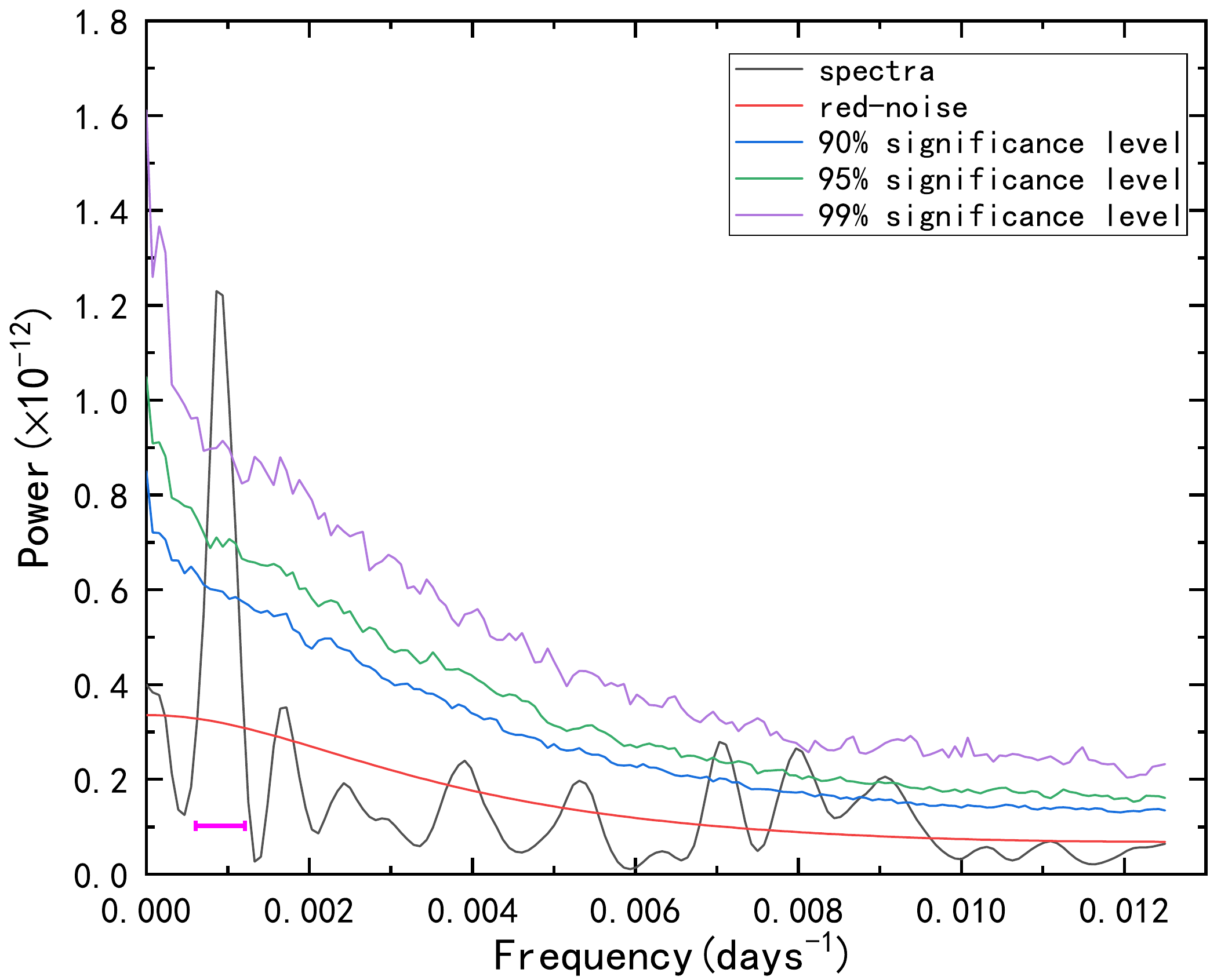}
	\caption{The power spectrum of the source in the \gr\ light curve (black solid line) and its confidence using the REDFIT method.
The red, blue, green, and purple curves represent the theoretical red-noise spectrum, 90\%, 95\%, and 99\% confidence levels, respectively. The magenta horizontal bar indicates 6-dB bandwidth.}
\label{Fig3}
\end{figure}

\begin{figure*}
\centering
\begin{minipage}[t]{0.47\textwidth}
\centering
\includegraphics[height=8.5cm,width=8.5cm]{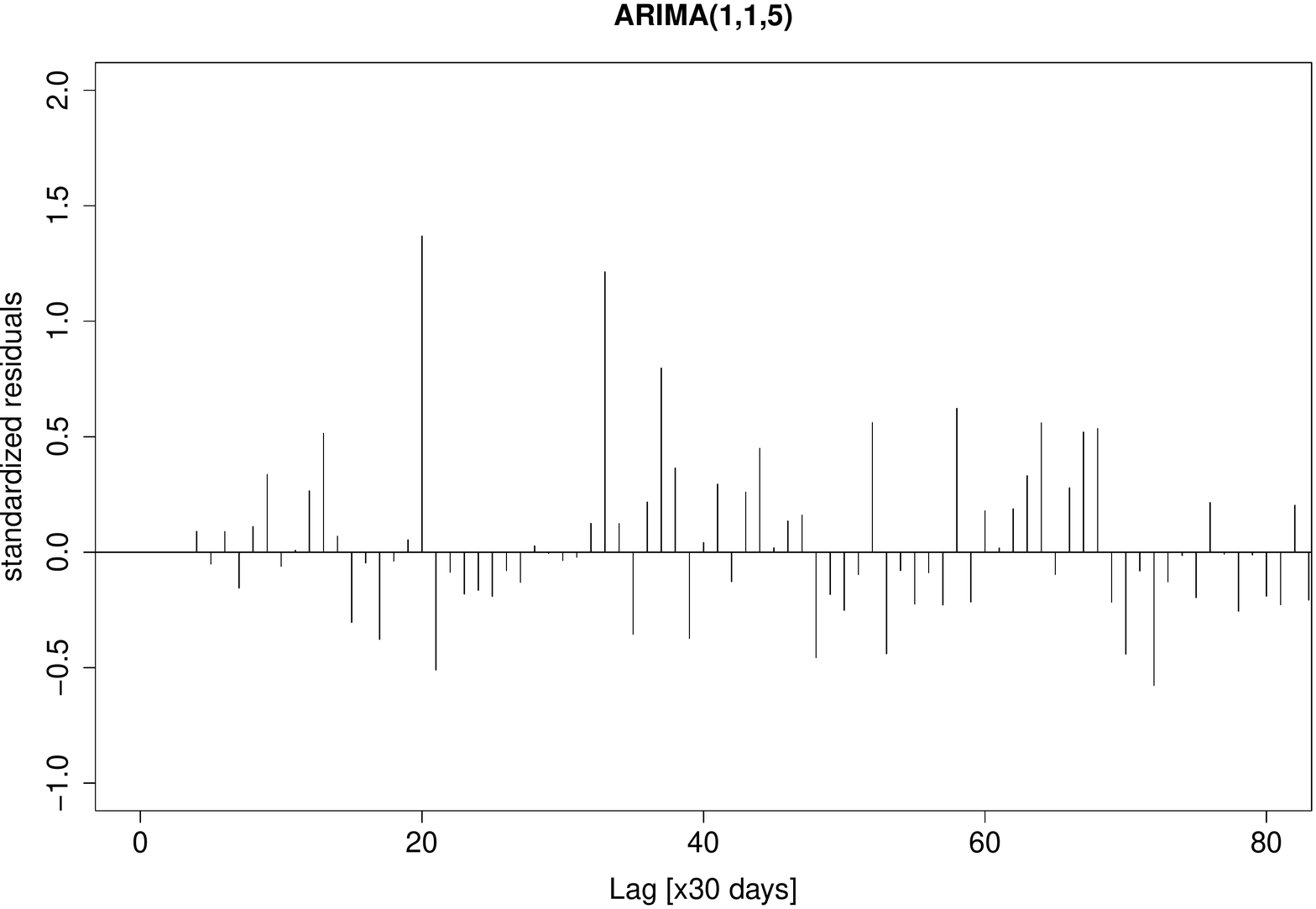}
\end{minipage}
\begin{minipage}[t]{0.47\textwidth}
\centering
\includegraphics[height=8.5cm,width=8.5cm]{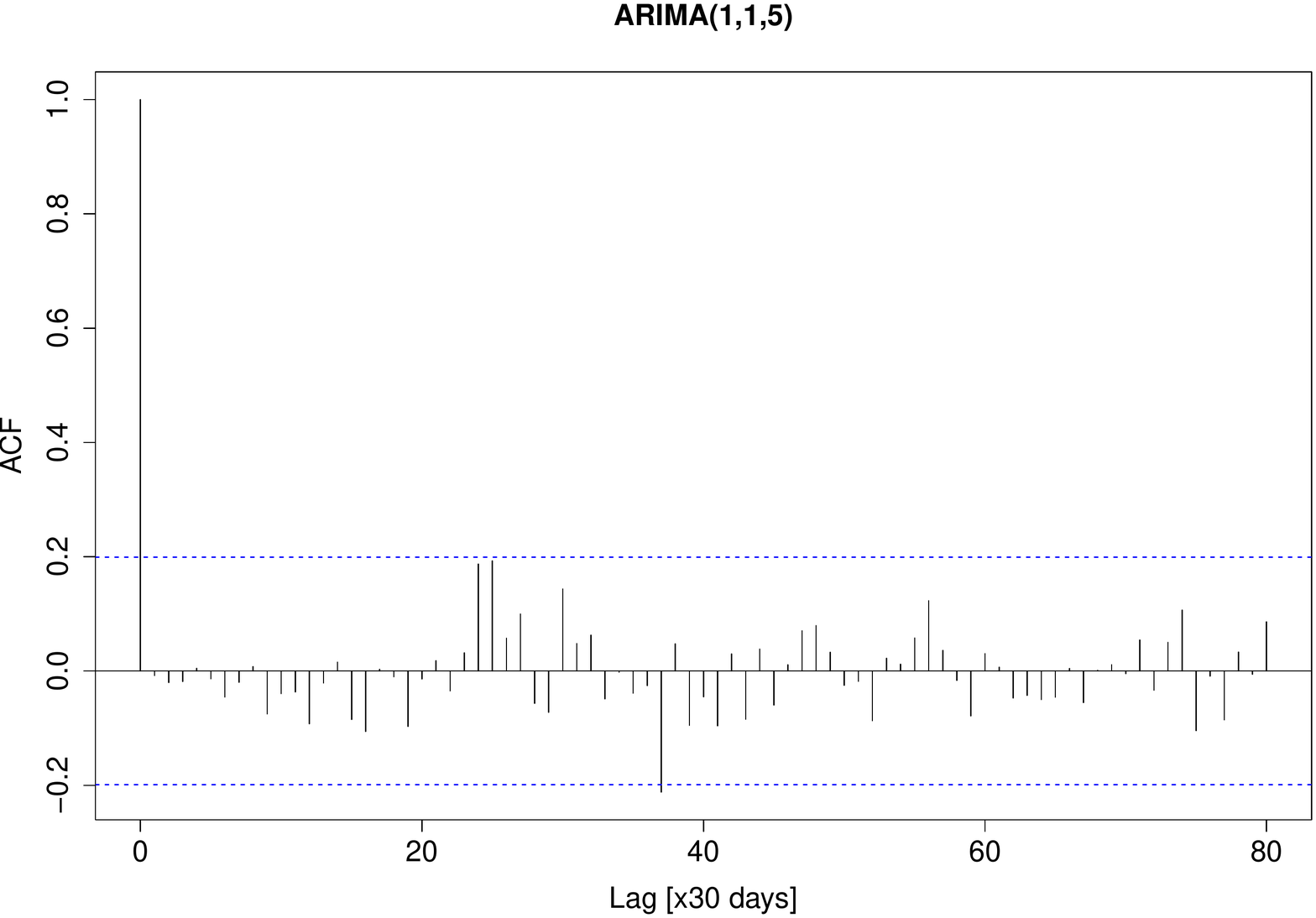}
\end{minipage}
\caption{Results of ARIMA model fitting for the 30 days bin $\gamma$-ray light curve. Left panel: standard residuals of the ARIMA(1, 1, 5) model fitting. Right panel: residuals ACF of the ARIMA(1, 1, 5) model fitting, where the blue dashed lines represent the 95$\%$ confidence interval.}
\label{Fig7}
\end{figure*}

\subsection{Fractional variability}
To further quantify and depicts the degree of variability in different bands, we estimated fractional variability amplitude $F_{\mathrm{var}}$ and, it can be expressed as \citep{2003MNRAS.345.1271V}
\begin{equation}
F_{\mathrm{var}} = \sqrt{\frac{S^2-\langle\sigma_{\mathrm{err}}^2\rangle}{\langle{x}\rangle^2}},
\end{equation}
where $S$, $\langle\sigma_{err}^2\rangle$ and $\langle{x}\rangle^2$ represent the standard deviation of the flux, the mean square error and the square of the average flux, respectively.
Nevertheless, the uncertainty of $F_{\mathrm{var}}$ is obtained with
\begin{equation}
\bigtriangleup{F_{\mathrm{var}}} = \sqrt{F_{\mathrm{var}}^2+err(\sigma_{\mathrm{N}}^2)}-F_{\mathrm{var}},
\end{equation}
where $err(\sigma_{\mathrm{N}}^2)$ is given by
\begin{equation}
err(\sigma_{\mathrm{N}}^2) = \sqrt{\left (\sqrt{\frac{2}{N}}\frac{\langle\sigma_{\mathrm{err}}^2\rangle}{\langle{x}\rangle^2}\right)^2
+\left (\sqrt{\frac{\langle\sigma_{\mathrm{err}}^2\rangle}{N}}\frac{2F_{\mathrm{var}}}{\langle{x}\rangle}\right)^2}
\end{equation}
$N$ is the number of data points set. Through the above three formulas, the fractional variability
amplitude values of \gr\ and G-band are $F_{\mathrm{\gamma}} = 0.59 \pm 0.02$ and $F_{\mathrm{G}} = 0.58 \pm 0.02$, respectively, which
indicates that there is relatively significant variability in this case.
Recently, \cite{2021MNRAS.506.1540L} reported that there might be a positive correlation between $F_{\mathrm{var}}$ and frequency, but this phenomenon is not apparent here.

\section{QPO behavior analysis and results}
\label{results}
In order to search for the possible periodicities and ascertain the corresponding significance,
the Weighted Wavelet Z-transform (WWZ), the Lomb-Scargle periodogram (LSP), and the
REDFIT were performed to analyze the \gr\ light curve. Moreover,
we also use the epoch folding method based on time domain analysis to search for periodicity.
By means of these methods, the reliability of a QPO signature can be verified.
We derive light curves in the \gr\ and optical bands from the data observed with Fermi-LAT and Gaia satellite.

\begin{figure*}
\centering
\includegraphics[width=0.65\textwidth]{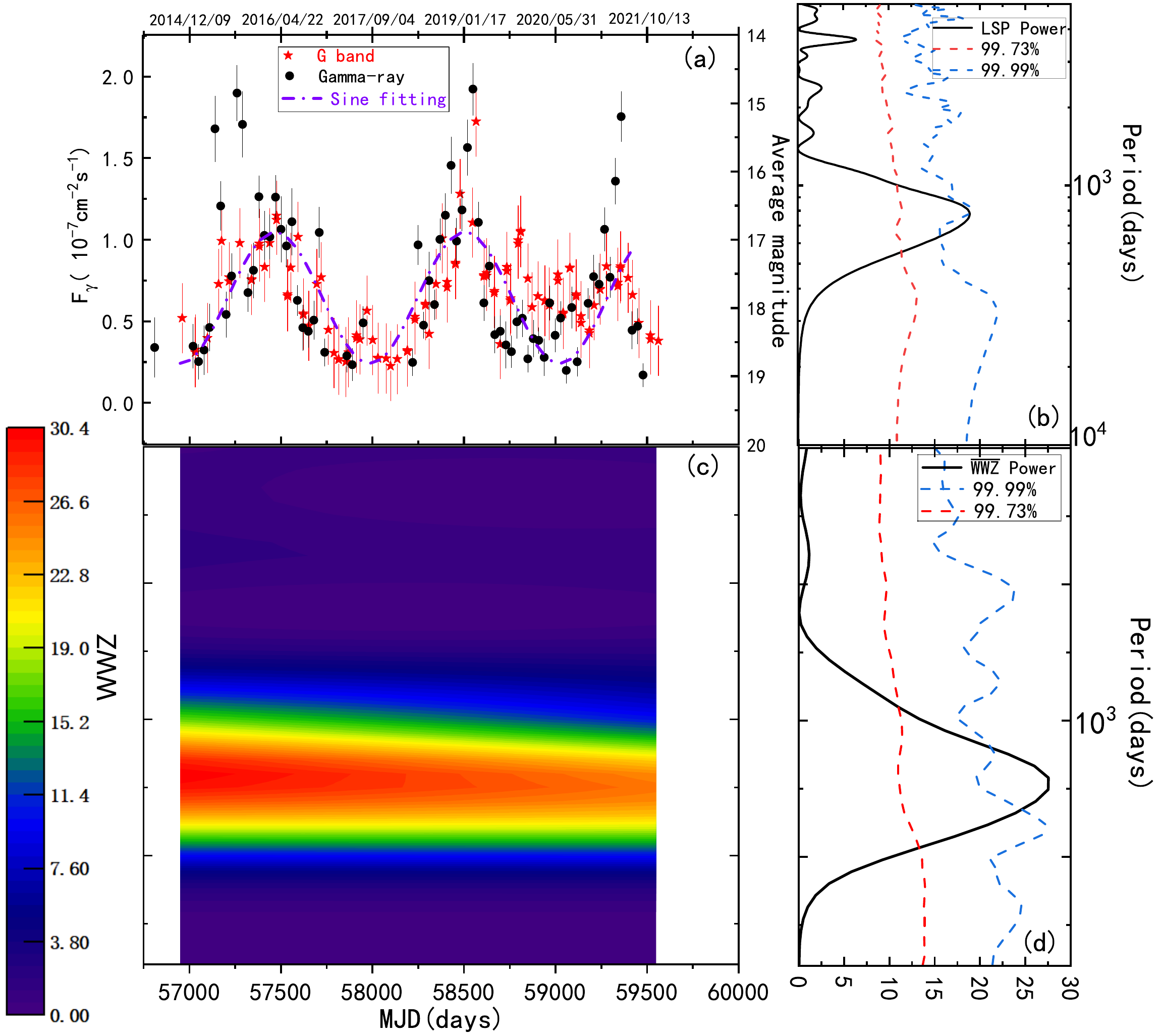}
%%\vspace{-20pt}\includegraphics[width=0.35\textwidth]{Fig3.pdf}
%%\plottwo{Fig1.pdf}{Fig3.pdf}
\caption{Panel (a): The red pentagram represents the light curve of the FSRQs PKS 0405-385 in the optical band from MJD 56960 to 59562.
The black dots represent the light curve of the source quasi simultaneous with the optical band.
Other Panels are the same as figure~\ref{Fig1}, but for the G-band light curve during MJD 56960 to 59562.}
\label{Fig4}
\end{figure*}

WWZ, a widely employed method of time series analysis in some astrophysical systems, was first proposed by \cite{Foster1996}.
This technique is a time (space) frequency localization analysis that gradually refines the signal (function) through expansion and translation operation.
Moreover, it can also automatically meet the requirements of time-frequency signal analysis and focus on any signal detail.
The collection of astronomical data is affected by the observation conditions resulting in the light curve being often non-equally spaced.
Then, \cite{Foster1996} proposed the idea of vector projection to deal with the problem that wavelet transform is very sensitive to the sampling interval of time series.
By implementing this idea, the WWZ power can be presented as a function of observing time and period, in which the peak in power represents the intensity and duration
of possible quasi-periodic behavior in the light curve.
The results of WWZ power show that the periodic modulation in \gr\ and optical bands is centered at 1025 $\pm$ 255 (2.8 $\pm$ 0.7 yr; the panel (d) of Figure~\ref{Fig1})
and 1296 $\pm$ 430 (3.6 $\pm$ 1.2 yr; the panel (d) of Figure~\ref{Fig4}) days with no significant changes over time, respectively.
We consider the half-width at half-maximum (HWHM) of the power peak fitted by the Gaussian function as the uncertainty of the quasi-periodic signal.

Lomb-Scargle periodogram (LSP), one of the most well-known methods of detecting periodicity, was first worked out by \cite{Lomb1976} and later ameliorated by \cite{Scarle1982}. The basic principle of the LSP method is to employ the least square method to fit the linear combination of a series of trigonometric functions ($y = a\cos\omega{t} + b\sin\omega{t}$).
Furthermore, the signal characteristics from the time domain to the frequency domain is converted.
The LSP power indicates a prominent peak around the timescale of 1002 $\pm$ 223 (2.7 $\pm$ 0.6 yr; the panel (b) of Figure~\ref{Fig1}) days in the \gr\ light curve.
A significant peak appears in the LSP power of G-band flux density at 1297 $\pm$ 425 (3.6 $\pm$ 1.2 yr) days, which is given in panel (b) of Figure~\ref{Fig4}.

The epoch folding method was also employed to prove the QPO signal further.
This method is mainly unaffected by the irregularity of observation data and the modulation shape of periodic components \citep{Bhatta2018}.
In Figure~\ref{Fig2}, one can see that this folded light curve varies with the phase, indicating substantial variability in the source brightness.
Nevertheless, the light curves of AGNs are mainly affected by red noise, which results from some stochastic processes in a jet plasma or the accretion disc \citep{Li2017}.
Emission from an AGN is usually autoregressive, so the first-order autoregressive (AR1) process can evaluate the red noise spectrum reasonably.
The calculation formula of the theoretical power spectrum of an AR1 process can be found in equation 2 of \citet{2020MNRAS.499..653K}.
We use the program REDFIT3.8e to complete the calculation, in which the parameters are set to $n_{50} = 2$ and a Hanning window is selected \citep{2002CG.....28..421S}. The resulting power spectrum (black) is shown along with the AR1 spectrum (red) in Fig~\ref{Fig3}. It can be seen that in the periodogram a distinct peak stands out around the timescale of 1086 $\pm$ 321 days with significance $\geqslant99\%$. It is worth noting that the REDFIT method can only give a maximum significance of 99 per cent. However, the significance of QPO detection based on the temporal spectrum is usually affected by the bandwidth penalty effect. In order to evaluate the impact of this effect on signal detection, the 6-dB bandwidth $B_w$ is commonly utilized: $B_w = \beta_w \cdot \bigtriangleup f$, where $\beta_w$ is the normalized bandwidth that depends on the spectral
window and $\bigtriangleup f$ is the fundamental frequency associated with $n_{50}$ \citep{1978IEEEP..66...51H,1997CG.....23..929S}. Considering different spectral windows (e.g., Rectangle, Triangle, Welch, Hanning, Blackman-Harris) and $n_{50}$ (0,1) values, we estimate the significance of QPO detection again. The overall results show that the quasi-periodic signal ($\geqslant 99\%$) may exist in the \gr\ light curve, in which the 6-dB bandwidth ranges from 0.00025 (only $2\%$ of the relevant frequency interval) to 0.00071.

In addition to the commonly used non-parametric Fourier-type (e.g., LSP) and wavelet analysis method, a statistical tool can also be employed to analyze the periodic characteristics in the light curve. Therefore, we fit the \gr\ light curve with autoregressive integrated moving average (ARIMA) model to further test whether the quasi-periodic behavior is consistent with a stochastic process \citep{1985JAWRA..21..721V,Wilson2016}. Using the Akaike Information Criterion (AIC), we selected the best-fit model from 72 ARIMA (p,q = 0...5, d = 0...1) models fitting the \gr\ light curve, which is the ARIMA (1,1,5) model corresponding to the minimum AIC value of 655. We can see from Fig~\ref{Fig7} that there is a spike at a lag of 1110 days that exceed the 95$\%$ confidence interval. This marginal evidence indicates that the quasi-periodic behavior of \gr\ light curve may be intrinsic.

Although we employ two independent techniques, LSP and WWZ, to search for periodicity in light curves, and the statistical properties of AGNs
light curves usually exhibit frequency dependent colored-noise-like behavior, likely to mimic a transient QPO behavior, especially in the low-frequency domain \citep{1978ComAp...7..103P,2003MNRAS.345.1271V,2016MNRAS.461.3145V}.
Moreover, spurious peaks might arise owing to other sampling effects including discrete sampling, finite observation period and uneven sampling of the light curve.
For this reason, it is essential to consider the significance estimation in QPO detection.
The periodogram of the source can usually be reasonably approximated as a power-law power spectral density (PSD) of the form $P(\nu) \propto \nu^{-\beta}$
where $\nu$ is the temporal frequency and $\beta$ the power-law index.
We performed a large number of simulations of light curves by randomizing both amplitude and phase to
construct the coloured noise background following the Monte Carlo method described in \citet{1995A&A...300..707T}.
To model the underlying red-noise PSD, we first estimate the power spectrum slope $\beta$ by fitting the log-periodogram of the LSP between 0.0001 and 0.01 $d^{-1}$
with a linear function following \citet{2005A&A...431..391V}.
We then generated $10^4$ artificial light curves to evaluate the significance of QPO against spurious detections in \gr\ and optical waveband based on the same
parameters as the original light curve, such as average value, standard deviation, and temporal baselines.
The results of the significance evaluation of QPO signals in \gr\ and optical bands are shown in Figure~\ref{Fig1} and Figure~\ref{Fig4}, respectively.
The red and blue dashed lines represent 99.73 per cent ($3\sigma$) and 99.99 per cent ($4\sigma$) confidence contour lines, respectively.
One can note that the significance of the observed periodic behavior at the period of $\sim$2.8 yr turned out to be $>4\sigma$ in high-energy \gr\
, and similarly the significance of the G-band power spectrum peak at the period $\sim$3.6 yr was evaluated to be $\sim4\sigma$.
Furthermore, the significant results of a similar periodic feature are presented in the WWZ analysis.

\section{Discussion and conclusions}
\label{sumdis}

\begin{figure}
\centering
\includegraphics[scale=0.3]{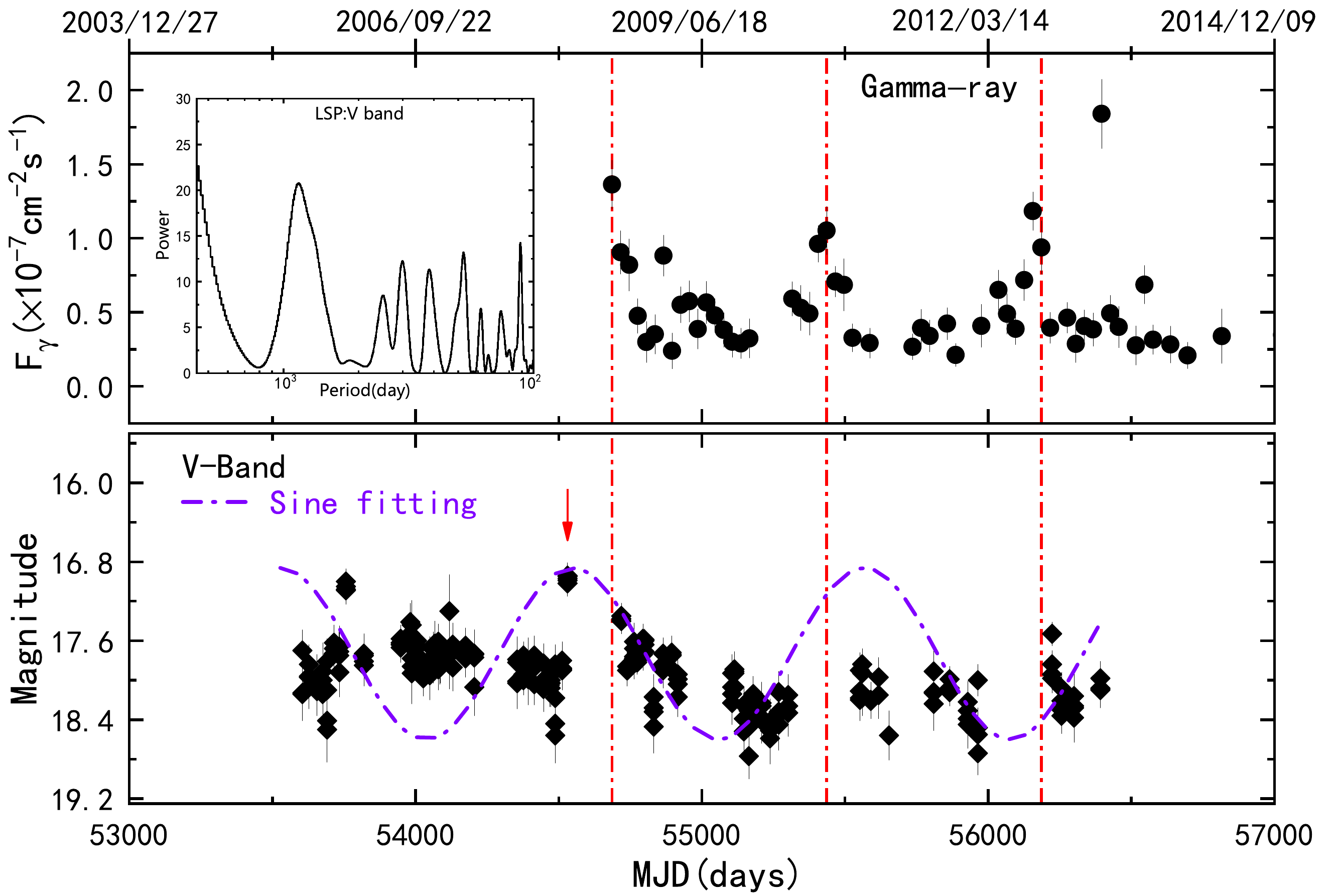}
	\caption{The red vertical dashed-dotted line indicates the similarity of flux density changes between the \gr\ and V band,
              in which a V-band LSP periodic diagram is embedded in the upper panel.
              The red arrow is the peak of the QPO cycle (MJD$\sim$54530). The purple dash-dotted line represents the sinusoidal fitting curve of the observation data.}
\label{Fig5}
\end{figure}

\begin{figure}
\centering
\includegraphics[height=6cm,width=7cm]{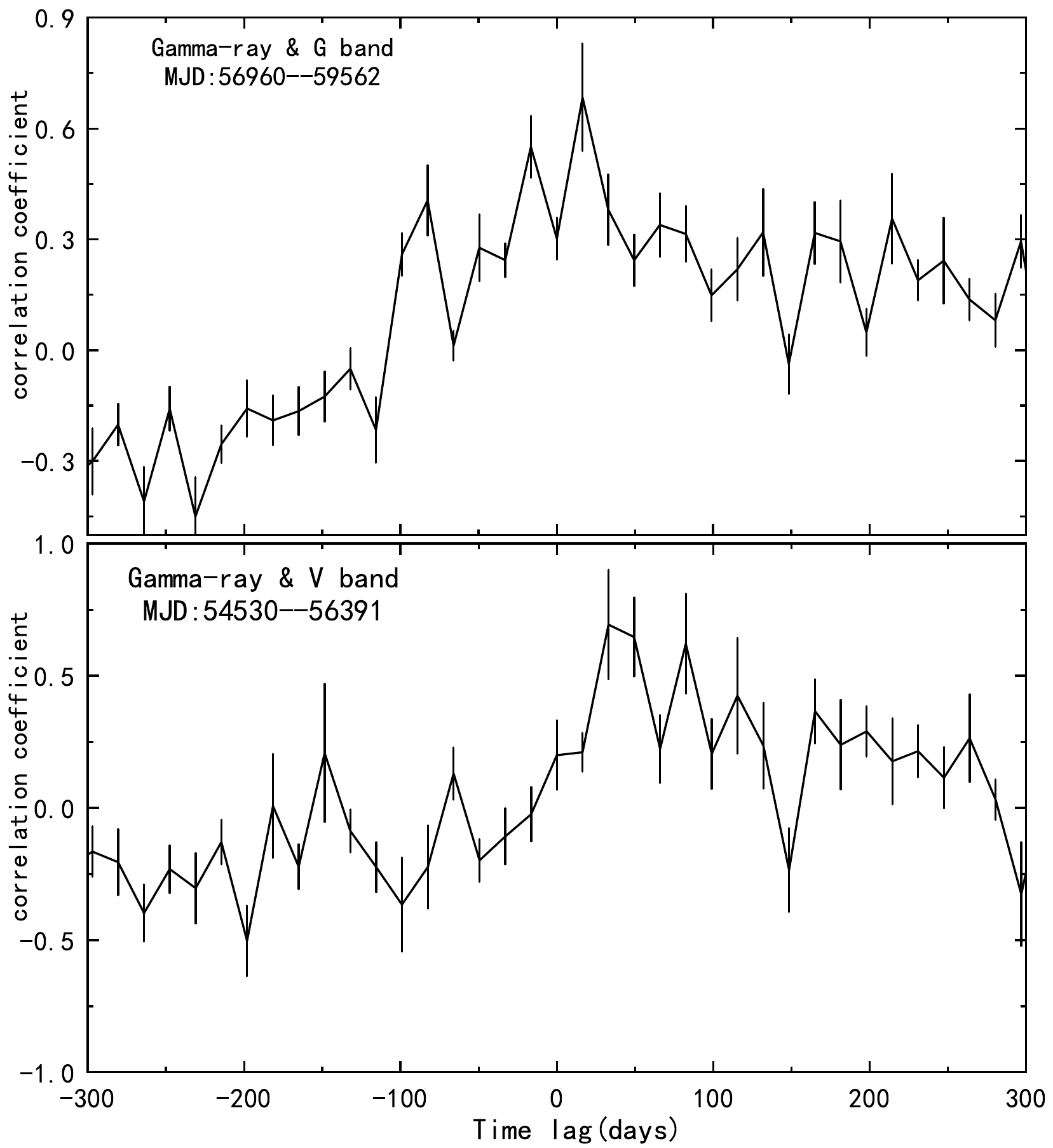}
	\caption{DCF results between \gr\ (time bin = 30 days) and optical band (G and V band),
        where the monitoring data in the same temporal coverage are used to perform DCF analysis.}
\label{Fig6}
\end{figure}

We dealt with and analyzed the long-term Fermi-LAT \gr\ observation data of blazar PKS 0405-385 during 2008-2021 ($\sim$13.2 yr),
and the results indicate that a highly probable quasi-periodic behavior of $\sim$2.8 years appears in the whole light curve.
In addition, the optical G-band data observed from the Gaia satellite during MJD 56960 to 59562 were also collected for this source.
Furthermore, we also employed LSP and WWZ methods to search for QPO behavior and then found that a conspicuous power spectrum peak appeared at $\sim3.6$ years.
Peaks at $T_{\gamma}\sim 1037 \pm 266$ days and $T_{\mathrm{opt}}\sim 1296 \pm 430$ days are identified at the same frequency within the errors in the two bands.
Since the temporal coverage of high-energy \gr\ is longer than that of the optical band, a periodic timescale of $\sim$2.8 years is adopted as the final result.
Interestingly, the V-band data from the Catalina Real-time Transient Survey during MJD 53604 to 56391 were also found, and the flux density change during MJD 54530 to 56391 was similar to that of $\gamma$-ray. We calculated flare time according to the cycle of 2.8 years and found that there was a prominent flare in the V-band at a time (MJD$\sim$54530, the red arrow) that matches the forecast (see Fig~\ref{Fig5}). Moreover, the upper plane of Fig~\ref{Fig5} shows a trace of $\sim$1000 days periodic feature obtained by LSP analysis.
This situation provides a shred of possible evidence that the QPO behavior should be longer than the \gr\ light curve duration.
We then evaluate the significance of potential QPO by modeling coloured noise as a simple power-law or as an auto-regression function of the first order.
The results suggest that a $\sim4.3\sigma$ confidence level is at the peak of the power spectrum in the \gr\ light curve
and a $4\sigma$ significance contour is near the peak of the G-band. Detection of the period with statistical significance depends strongly on the steepness of the red noise PSD slope $\beta$. Generally, the significance detection at the quasi-periodic signal is inversely proportional to the steepness of the PSD slope \citep{2021MNRAS.508.3975K}. Therefore, we also use the standard theory of linear regression to evaluate the slope uncertainty based on the least square method with law, $err(\beta) = \sqrt{n\sigma^2/\bigtriangleup} \approx 0.26$, where $n$ is the number of frequencies used in the fitting, $\bigtriangleup = n\sum_{j=1}^n \log(f_j)^2 - (\sum_{j=1}^n \log(f_j))^2$, and $\sigma^2 = \pi^2/6(\ln(10))^2$ is the variance of the log-periodogram ordinates about the spectrum \citep{2005A&A...431..391V}. We take $err(\beta)$ as the upper and lower limits of the PSD slope to reevaluate the significance of QPO behavior, and the results indicate that the confidence level of the QPO signal ranges from $\sim4\sigma$ to $\sim4.7\sigma$. In view of the similarity of flux density variation between \gr\ and optical bands (G and V band),
we analyzed the cross-correlation of the \gr\ to optical flux using discrete correlation function,
followed the prescription described by \citet{1988ApJ...333..646E} and presented it in Fig~\ref{Fig6}.
There is an obvious main peak at an almost null delay, and this strong gamma-ray-optical
correlation is expected by the leptonic single-zone model of blazar emission \citep{2014ApJ...797..137C}.

Although the QPO characteristics of the emissions from AGNs are still controversial, several explanations for the origin of quasi-periodic behavior have been widely discussed.
Several models based on fluctuations or oscillations in the accretion disk have been proposed to explain that the timescale of QPO
variability is intra-day variability (IDV) from several minutes to hours, especially in X-ray binaries \citep{Remillard2006}.
A crucial case is that a 1-hour X-ray modulation is reported in RE J1034+396, and a mass range of 4$\times 10^5$--10$^7$\,$M_{\odot}$ is evaluated \citep{Gierlinski2008}.
The simplest explanation for such nearly IDV might be that the flux arises from extreme orbiting hot spots on the disks at, or close to,
the innermost stable circular orbit around black holes allowed by general relativity (\citealt{1993ApJ...406..420M,2009A&A...506L..17L}, and references therein).
Using such a simple model, \citet{2009A&A...506L..17L} gave a black hole (BH) mass range of 3.29$\times 10^7$--2.09$\times10^8$\,$M_{\odot}$
in PKS 2155-304 with a quasi period of 4.6 hours.
However, \citet{2002ApJ...581..103R} analyzed the periodicity of PKS 0405-385 from June 8 to 10, 1996 using Lomb method, and found a radio QPO of 2.5 hours in the 8.6 GHz light curve.
In the simplest model, the black hole mass of the target can be evaluated with law, $M/M_{\odot} = 3.23 \times 10^4P/((r^{3/2}+a)(1+z))$, where $P$ is the observation period in seconds,
$a$ is the angular momentum parameter, $z = 1.285$ is the redshift and $r = R/R_{\mathrm{g}}$ \citep{2009ApJ...690..216G}.
The nominal BH mass of PKS 0405-385 with a radio QPO of 2.5 hours is estimated to be $8.7 \times 10^6 M_{\odot}$ for a non-rotating BH (with $r = 6$ and $a = 0$)
and $5.5 \times 10^7 M_{\odot}$ for a maximally rotating BH (with $r = 1.2$ and $a = 0.9982$).
It is expected that IDV can be detected in other bands for this case.
Other possible mechanisms for generating observed QPO are related to a disk or relativistic jets, such as small epicyclic deviations from exact planar motions within a thin accretion disk, magnetically choked accretion flows, and pulsational accretion flow instabilities \citep{2005AN....326..782A,2008ApJ...679..182E,2011MNRAS.418L..79T}.
However, the $\sim$2.8 yr QPO we found in the \gr\ light curve is obviously different from the time scales mentioned in the above scenarios.

Alternatively, QPO-like flux modulations might be related to a geometrical scenario with the relativistic motion of the enhanced emission (or blobs) along the helical path of a jet.
This scenario has been invoked to explain QPOs in PKS 2247-131 \citep{Zhou2018}, 3C 454.3 \citep{2021MNRAS.501...50S}, and OT 081 \citep{2021MNRAS.506.1540L} recently. When the emitting blob's moving helically within the magnetized jet with a high bulk Lorentz factor ($\Gamma$),
due to the relativistic effects, the periodic changes in the viewing angle $\theta$ cause the Doppler boosted emission to be periodically modulated. The viewing angle of an emitting blob's motion depends largely on the observed period, pitch angle $\phi$ between the emitting blob's motion and the jet's axis,
and the inclination angle $\psi$ of the jet with respect to the observers \citep{2017MNRAS.465..161S}.
By investigating the relationship between $\Gamma$ and jet angle, \citet{Bhatta2018} gave the viewing angle of a typical blazar in the range of 1-$5^\circ$, and pointed out that slight changes in viewing angle ($\sim1.5^\circ$) are sufficient to improve the observed brightness.
Due to Doppler boosting, the observed period $P_{\mathrm{obs}}$ should be much smaller than that of the rest frame $P_{\mathrm{rest}}$ at the host galaxy: $P_{\mathrm{obs}} = (1-\beta \; \cos \phi \; \cos \psi)P_{\mathrm{rest}}$.
Then, the distance that the blob moves in six cycles would be approximately $D = 6c\beta \, P_{\mathrm{rest}} \; \cos \phi \; \sin \psi \approx 25.7$ pc.
Here, we use the typical parameters in blazars: $\phi = 2^\circ$, $\psi = 3^\circ$, and $\Gamma = 10$ \citep{1997ApJ...490L...9K,Zhou2018}.
However, the Doppler factor depends upon viewing angle, $\theta$, and the velocity of the shock propagating down the jet, $\upsilon_{\mathrm{jet}}$, as $\delta = [\Gamma(1-\beta \; \cos \theta)]^{-1}$,
where $\beta = \upsilon_{\mathrm{jet}}/c$ and $\Gamma = (1-\beta^2)^{-1/2}$. When the moving blob in the jet dissipates,
the quasi-periodic variability of flux may become less obvious or even disappear, which naturally explains the transient nature of the QPOs behavior.
In addition, the observed QPO may also be caused by Lense-Thirring precession of the inner portions of the accretion disc,
but its origin is unlikely to produce modulation for more than a few months \citep{Stella1998,2021MNRAS.501.5997T}.

Finally, we also consider a supermassive black hole binary (SMBHB) system to explain the long-term periodic temporal signals.
\citet{1988ApJ...325..628S} ascribed the optical QPO in the blazar OJ 287 to this mechanism.
Several other discovered candidate QPOs sources are also discussed in this model, such as $\sim$2.18 yr in PG 1553+113 \citep{2015ApJ...813L..41A}, $\sim$2.1 yr in PKS 0301-243 \citep{Zhang2017c}, $\sim$3.0 yr in 3C 66A \citep{2020MNRAS.492.5524O}, and $\sim$4.69 yr in PKS J2134-0153 \citep{2021MNRAS.506.3791R}.
The Keplerian orbital motion of an SMBHB would trigger periodic accretion perturbations, or the gravitational torque from a
companion would induce the jet-precessional and nutational motions in misaligned disc orbits
and yield periodic timescales in the range of $\sim$1 to $\sim$25 years \citep[][and reference therein]{1997ApJ...478..527K,2004ApJ...615L...5R,2006MmSAI..77..733K,2017ApJ...836..220C}.
The observed period $P_{\mathrm{obs}}$ of $\sim$2.8 yr is corrected to the intrinsic orbital period $P_{\mathrm{int}} = P_{\mathrm{obs}}/(1+z) \simeq 793$ days for the cosmological redshift.
According to \citet{2015PASP..127....1L}, the mass of the primary black hole can be estimated via the relation $M \simeq (\Gamma^2P_{\mathrm{int}})^{8/5}R^{3/5}M_{\odot}^6$, where $R$ is the mass ratio between the primary and secondary components and $M_{\odot}^6$ is $10^6M_{\odot}$.
The value range of R is generally 0.01-0.1, where R = 0.1 is used to calculate the mass of the primary black hole $M \simeq 10^{8.7}M_{\odot}$ \citep{1980Natur.287..307B}.
Based on the value of $M$, we can calculate the separation between two black holes as $r \sim 123 r_{\mathrm{s}} \sim 0.008$ parsec, where $r_{\mathrm{s}}$ represents the Schwarzschild radius of the primary black hole \citep{2010RAA....10.1100F}. Such a milliparsec separation might be too small to yield an observable orbital decay time-scale in the evolution of SMBHB systems.
The orbital decay timescale in the gravitational waves (GW) driven regime can be estimated with $\tau \sim 3.05\times10^{-6}(M/10^3M_{\odot}^6)(r/r_{\mathrm{s}})^4$yr \citep{Bhatta2018}.
If the quasi-periodic behavior originates from such a system, we may predict that the system will experience gravitational coalescence within 56 centuries accompanied with the emission of GW. Therefore, this object may be one of the future potential targets of GW detector observation.

\section*{Acknowledgements}
{
This research or product makes use of public data provided by ESA/Gaia/DPAC/CU5.
JF is partially supported by National Natural Science
Foundation of China (NSFC) under grants  11873042, U2031107, the Program of Yunnan University (WX069051, 2017YDYQ01), the grant from Yunnan Province (YNWR-QNBJ-2018-049)
and the National Key R\&D Program of China under grant No.2018YFA0404204).
}

\end{document}